\global\def\draftcontrol{0}

   \def\versionno{ superpotential }

\catcode`\@=11

\newcommand\makepapertitle{\par

  \begingroup
    \renewcommand\thefootnote{\@fnsymbol\c@footnote}%
 \newpage
     \global\@topnum\z@   
     \@makepapertitle
     \thispagestyle{empty}\@thanks
  \endgroup
  \setcounter{footnote}{0}%
  \global\let\thanks\relax
  \global\let\makepapertitle\relax
  \global\let\@makepapertitle\relax
  \global\let\@thanks\@empty
  \global\let\@author\@empty
  \global\let\@date\@empty
  \global\let\@title\@empty
  \global\let\title\relax
  \global\let\author\relax
  \global\let\date\relax
  \global\let\and\relax
  \def\version{\let\version\@version\@gobble}
}
\def\@makepapertitle{%
  \newpage
   \ifnum\draftcontrol=1 {}
   \version\versionno
   \vskip 5.5em%
   \else
   \hfill\hbox to 3.5cm {\parbox{5cm}{\@pubnum}\hss}%
   \vskip 6.5em%
   \fi
   \begin{center}%
   \let \footnote \thanks
      {\hskip -0\textwidth \hbox to 1\textwidth%
        {\centerline{\Large\bf{\noindent%
    \parbox[t]{1.3\textwidth}{\begin{center}\@title\end{center}}}}}}%
     \vskip 1.5em%
     {\normalsize
       \lineskip .5em%
       \begin{tabular}[t]{c}%
         \@author
       \end{tabular}\par}%
     \vskip 1.5em%
     {\@bstract}%
     \end{center}%
     \vfill
     \@date%
     \vskip 1.5em%
   \par
}

\gdef\@pubnum{}
\def\pubnum#1{%
  \gdef\@pubnum{#1}}

\gdef\@bstract{}
\def\Abstract#1{%
  \gdef\@bstract{%
   \parbox{\textwidth-0pc}{%
   \centerline{\bf Abstract}\penalty1000
   \noindent
   \renewcommand\baselinestretch{1.0}
   {#1}}}
}

\gdef\@email{}
\def\email#1{%
   \gdef\@email{%
   Email: {\tt #1}}
}

\def\ps@paper{\let\@mkboth\@gobbletwo%
     \ifnum\draftcontrol=1
        \def\@oddfoot{\hbox to \textwidth{\tiny \versionno \hfil\tiny\draftdate}%
        \hskip -\textwidth \hbox to \textwidth{\hfil\rm\thepage\hfil}}%
     \else\def\@oddfoot{\hbox to \textwidth{\hfil\rm\thepage\hfil}}
     \fi
     \let\@evenfoot\@oddfoot
}

\def\body{\clearpage
          \pagestyle{paper}
        }

\def\@version#1{\ifnum\draftcontrol=1
\typeout{}\typeout{#1}\typeout{}
\vskip3mm\centerline{\hbox{\fbox{\normalsize{\tt DRAFT -- #1 -- }
                   {\draftdate}}}}\vskip3mm
\fi}
\let\version\@version
\long\def\eqlabel#1{\ifnum\draftcontrol=1
                    \tag@false  
                    \tag*{(\theequation) \hbox to -0.2cm{\hspace{0cm}\small{#1}\hss}}
                    \refstepcounter{equation}
                    \edef\@currentlabel{\theequation}
                    \ltx@label{#1}
                    \else
                    \label{#1}
                    \fi
                    }
\let\st@bibitem\@bibitem
\let\st@lbibitem\@lbibitem
\ifnum\draftcontrol=1
  \def\@bibitem#1{%
    \st@bibitem{#1}\a@@label{#1}\ignorespaces}
  \def\@lbibitem[#1]#2{%
    \st@lbibitem[#1]{#2}\a@@label{#2}\ignorespaces}
  \def\a@@label#1{%
    \gdef\a@lab{\smash{\normalfont\small#1}}
    \ifvmode
      \if@inlabel
        \global\setbox\@labels\hbox{%
          \llap{\a@lab\let\a@lab\relax
                \kern\@totalleftmargin\kern\marginparsep}%
          \box\@labels}%
      \fi
    \fi}
\fi

\documentclass[12pt,letterpaper]{article}
\usepackage{amsmath}
\usepackage{amsmath}
\usepackage{amsmath}
\usepackage{amssymb}
\usepackage{amssymb}
\usepackage{amssymb}
\usepackage{amsmath,bm,amsfonts,amssymb,array,calc,amsthm,rotating,cite}
\usepackage{epsfig,psfrag}
\usepackage{graphicx}
\usepackage{color}
\usepackage[colorlinks=true]{hyperref}
\usepackage[all]{xy}

\renewcommand\baselinestretch{1.25}
\renewcommand\section{\@startsection {section}{1}{\z@}%
                                   {-3.5ex \@plus -1ex \@minus -.2ex}%
                                   {2.3ex \@plus.2ex}%
                                   {\normalfont\large\bfseries}}
\renewcommand\subsection{\@startsection{subsection}{2}{\z@}%
                                   {-3.25ex\@plus -1ex \@minus -.2ex}%
                                   {1.5ex \@plus .2ex}%
                                   {\normalfont\normalsize\bfseries}}
\renewcommand\subsubsection{\@startsection{subsubsection}{3}{\z@}%
                                   {-3.25ex\@plus -1ex \@minus -.2ex}%
                                   {1.5ex \@plus .2ex}%
                                   {\normalfont\normalsize\it}}
\renewcommand\paragraph{\@startsection{paragraph}{4}{\z@}%
                                   {-3.25ex\@plus -1ex \@minus -.2ex}%
                                   {1.5ex \@plus .2ex}%
                                   {\normalfont\normalsize\bf}}
\renewcommand\subparagraph{\@startsection{subparagraph}{5}{\z@}%
                                   {-1.25ex\@plus -1ex \@minus -.2ex}%
                                   {0ex \@plus .2ex}%
                                   {\normalfont\normalsize\it}}

\numberwithin{equation}{section}

\renewcommand*\l@section[2]{%
  \ifnum \c@tocdepth >\z@
    \addpenalty\@secpenalty
    \addvspace{.5em \@plus\p@}%
    \setlength\@tempdima{1.5em}%
    \begingroup
      \parindent \z@ \rightskip \@pnumwidth
      \parfillskip -\@pnumwidth
      \leavevmode \bfseries
      \advance\leftskip\@tempdima
      \hskip -\leftskip
      #1\nobreak\hfil \nobreak\hb@xt@\@pnumwidth{\hss #2}\par
    \endgroup
  \fi}
\renewcommand*\l@subsection{\addvspace{.0em \@plus\p@}\@dottedtocline{2}{1.5em}{2.3em}}
\renewcommand*\l@subsubsection{\addvspace{-.2em \@plus\p@}\@dottedtocline{3}{3.8em}{3.2em}}

\definecolor{refcol}{rgb}{0.0,0.0,0.2}
\definecolor{eqcol}{rgb}{.2,0,0}
\definecolor{purple}{cmyk}{0,1,0,0}

\gdef\@citecolor{refcol} \gdef\@linkcolor{eqcol}
\gdef\@urlcolor{refcol}
\def\colorlinkspurple{\gdef\@urlcolor{purple}}
\def\colorlinksblue{\gdef\@urlcolor{blue}}
\def\colorlinksred{\gdef\@urlcolor{red}}

\def\revise#1       {\raisebox{-0em}{\rule{3pt}{1em}}%
                     \marginpar{\raisebox{.5em}{\vrule width3pt\
                     \vrule width0pt height 0pt depth0.5em
                     \hbox to 0cm{\hspace{0cm}{%
                     \parbox[t]{4em}{\raggedright\footnotesize{#1}}}\hss}}}}

\catcode`\@=12
\newcommand{\beq}{\begin{eqnarray}}
\newcommand{\eeq}{\end{eqnarray}}
\begin{document}

\title{
Mirror Symmetry, D-brane Superpotential and Ooguri-Vafa Invariants of Compact Calabi-Yau Manifolds}

\author{
Shan-Shan Zhang,~~~Fu-Zhong Yang\footnote{Corresponding author~~~
E-mail: fzyang@gucas.ac.cn} \\[0.2cm]
\it College of Physical Sciences, Graduate University of Chinese Academy of Sciences\\
\it   YuQuan Road 19A, Beijing 100049, China}

\Abstract{~~~~~The D-brane superpotential is very important in the low energy effective theory. As the generating function of all disk instantons from the worldsheet point of view, it plays a crucial role in deriving some important properties of the compact Calabi-Yau manifolds. By using the GKZ-generalized hypergeometric system, we will calculate the B-brane superpotentials of two non-fermat type compact Calabi-Yau hypersurfaces in toric varieties,respectively. Then according to the mirror symmetry, we obtain the A-model superpotentials and the Ooguri-Vafa invariants for the mirror Calabi-Yau manifolds.}

\makepapertitle

\body

\version\versionno

\vskip 1em

\newpage

\section{Introduction}

~~~~The theory of topological string which is derived from the two dimensional $(N,\hat{N})=(2,2)$ superconformal field theory has gained a great development over the past years and it has influenced deeply on the mathematics. The D-brane superpotential, the generating functional of correlation function, is particularly vital physical quantity which is a section of a special holomorphic line bundles of the moduli space from the mathematical perspective. Through the superpotential we can derive a series of important properties for the CY manifolds such as Yukuwa couplings, Ooguri-Vafa invariants and so on. Therefore the calculation of the superpotential is very meaningful.

Some important properts of the moduli spaces for various Calabi-Yau
manifolds \cite{Yau:9406055, Berglund:9506091, Alim:1010.0977} has
been well studied via the mirror symmetry which was first mentioned
in the local operator algebra of the $N=2$ string theory
\cite{Lerche:1990}. It is well known that mirror symmetry connects
two different moduli spaces which are respectively parameterized by kahler
geometry deformation and complex geometry deformation in A- and B-model. In A-model there exist contributions from the instantons  while there is none in B-model. So calculating superpotential directly in A-model is considerably difficult. In fact only in several special cases we know the corresponding brane configuration on mirror A-model side for a given brane configuration in the compact CY manifold \cite{Jockers:0808.0761} derived from the GKZ system in B-model. In GKZ system the superpotential is related to period integral. And the Hodge theoretic approach \cite{Yau:9511001} provides a
useful insight on studying the period integrals for CY manifolds period integral satisfies Picard-Fuchs differential equation
which is closely related to the GKZ system..

Recently, for compact CY manifolds, there are some great development in calculating the quantum corrected domain wall tensions on the CY threefolds via open-closed mirror symmetry \cite{Walcher:0605162, Morrison:0709.4028}.
The properties of some compact Calabi-Yau
manifolds has been studied in refs. \cite{JA, JM, KGS}.
In this note, we compute the D-brane superpotentials for two
non-fermat CY threefolds in detail via mirror maps and GKZ
hypergeometric system.

The structure of this paper is as follows. In section 2 we describe
the generalized GKZ hypergeometic system. The solution of GKZ hypergeometic system is
just the integral period. We also outline the approach to constructing the corresponding
polyhedron $\Delta$ and its mirror polyhedron $\Delta^*$ for the
Calabi-Yau manifold. Then we review how to calculate
superpotential. In section 3 we analyze two non-fermat type compact CY manifolds in toric varietis, respectively. and compute their superpotentials
as well as some disk invariants with the method referred previously. The last section is
the conclusion.

\section{Toric Geometry, Relative Period Integrals and GKZ System}
~~~~We divide this section into two parts to review some related background.

\subsection{Superpotential on D-brane}

 ~~~~In the presence of some background fluxes and space-filling D-branes, the type II
string theory compactification on Calabi-Yau manifold gives rise to the $N=1$ low effective energy
 theories \cite{JA}, whose effective superpotential is captured by the relative period integral $¡ª¡ª$ of the holomorphic three form $\Omega(z)$
around the integral relative cycle with boundaries in the D-branes
\cite{Fuji:1011.2347}. As is listed in the refs.
\cite{Baumgartl:0704.2666, Witten:9207094, Witten:9706109,
Vafa:0012041} that the above integral is derived from the action of
a holomorphic Chern-Simons theory on the brane which wraps the
holomorphic curves.

For a $D$-brane wrapping internal cycles of Calabi-Yau manifold $X$,
the corresponding effective superpotential is \cite{Witten:9207094}
\beq \mathcal{W}_{brane} = \int_{X}\Omega\wedge Tr[A \wedge
\bar{\partial}A + \frac{2}{3}A\wedge A\wedge A] \eeq where if there are $N$ branes, $A$ is a
holomorphic $U(N)$ gauge connection on $X$
and $\Omega$ is the holomorphic three form on $X$. For type IIB
string, the effective superpotential is a linear combination of the relative period integrals
\cite{Vafa:0012041, Vafa:0105045, Mayr:0108229}. \beq
\mathcal{W}_{brane}(\varphi, \xi) =
\hat{N}_{a}\hat{\Pi}^{a}(\varphi, \xi) \eeq where $\hat{N}_{a}$
stands for the homology class, which is wrapped
with the D-brane. $\hat{\Pi}^{a}(\varphi, \xi)$ represents the
period integral \beq \hat{\Pi}^{a}(\varphi,
\xi) = \int_{\gamma^{a}(\xi)}\Omega(\varphi) \eeq Here, $\xi$ and
$\varphi$ stands for the open- and closed-string
moduli respectively.  The internal background fluxes $H = H^{RR} + H^{NS}$
lead to an effective superpotential \cite{E, I, J,JA,
JM, Vafa:9906070, KGS} which is defined by \beq \mathcal{W}_{Flux} = \int \Omega\wedge H = \int
\Omega\wedge(H^{RR} + \tau H^{NS}) \eeq where $\Omega$ denote the
holomorphic three-form on the Calabi-Yau manifold, and $\tau$ denote
the complex couplings for the type II string of B-model. In this
note, wo only consider the RR flux, the induced superpotential
becomes
\beq \mathcal{W}_{flux} = \int \Omega\wedge H^{RR} =
\sum_{\alpha} N_{\alpha} \Pi^{\alpha}(\varphi)
\eeq
Therefore the combined superpotential generated by D-brane and flux is
\cite{Lerche:0207259, Lerche:0208039}
\beq \mathcal{W}(\varphi, \xi)
= \mathcal{W}_{brane}(\varphi, \xi) + \mathcal{W}_{flux}(\varphi) =
\sum N_{\Sigma}\Pi_{\Sigma}(\varphi, \xi)
\eeq
here the coefficient $N_{\Sigma}$ denotes both the D-brane topological charge and the RR
flux quantum data and $\Pi_{\Sigma}(\varphi, \xi)$ denotes the
integral of the three-form $\Omega(\varphi)$ over the three-chains
in the relative integer homology group, which is defined by
\beq
\Pi_{\Sigma}(\varphi, \xi) = \int_{\Gamma^{\alpha}(\xi)}
\Omega(\varphi)~~ ,~~~ \Gamma^{\alpha}(\xi) \in H_{3}(Y, S, Z)
\eeq
The relative period integral referred previously
$\hat{\Pi}^{a}(\varphi, \xi)$ is equal to the domain wall tension
$\mathcal{T}(\varphi, \xi)$  \cite{Vafa:0012041,
Lerche:0207259, Lerche:0208039, Jockers:0808.0761, Witten:9706109}.
$\mathcal{T}(\varphi, \xi)$ is defined as
\beq \mathcal{T}(\varphi,
\xi) = \mathcal{W}(C_{(\varphi, \xi)}^{+}) -
\mathcal{W}(C_{(\varphi, \xi)}^{-})
\eeq
At its critical point $\xi = z$, $\mathcal{T}(\varphi, \xi)$ is
identical to the on-shell domain wall tension $T =
W(C_{\varphi}^{+}) - W(C_{\varphi}^{-})$. As is depicted in the
refs. \cite{Morrison:0709.4028, Yau:0910.4215, Alim:1010.0977,
Clemens:0206219}, at the critical points, the domain
wall tensions are considered as normal function from which the
Abel-Jacobi invariants can be derived .

For the D-brane in A-model, the superpotential which is expressed in terms
 of the flat closed/open coordinates can be calculated as the
generating functional of the correlation functions
\cite{Baumgartl:1007.2447, Vafa:0012041, Lerche:0207259,
Alim:0901.2937, Ooguri:9912123}.It is defined by
\beq \mathcal{W}(t,
\hat{t}) = \sum_{\vec{k}, \vec{m}}G_{\vec{k}, \vec{m}} q^{d
\vec{k}}\hat{q}^{d \vec{m}} = \sum_{\vec{k},
\vec{m}}\sum_{d}n_{\vec{k}, \vec{m}}\frac{q^{d \vec{k}} \hat{q}^{d
\vec{m}}}{k^2}
\eeq
Here, $q = e^{2\pi it}$, $\hat{q} = e^{2\pi i\hat{t}}$ and $n_{\vec{k}, \vec{m}}$ is the Ooguri-Vafa invariant.
Mirror symmetry, which indicates that the two superpotentials for D-branes
in A- and B-model, respectively, are related to each other by the mirroe map,  gives us a method
to computing the Ooguri-Vafa invariant which is closely related to
the open Gromov-Witten invariant $G_{\vec{k}, \vec{m}}$
\cite{Jockers:0808.0761}.
The  superpotentials and the Ooguri-Vafa invariants are  as follows
\cite{Walcher:0605162, TT} :
\beq
\frac{W^{(\pm)}(z(q))}{\omega_{0}(z(q))} = \frac{1}{(2\pi i)^{2}}
\sum_{k\in odd} \sum_{d_{1}\geq0,d_{2}\in odd}
n_{d_{1},d_{2}}^{(\pm)} \frac{q_{1}^{kd_{1}}
q_{2}^{kd_{2}/2}}{k^{2}} \eeq where $q_{a} = e^{t_{a}}~~(a = 1, 2)$
and \beq \omega_{0}(z) = \sum_{n} c(n) z^{n} = \sum_{n}
\frac{\prod_{j} \Gamma(\sum_{a} l_{0j}^{(a)} n_{a} + 1)}{\prod_{i}
\Gamma(\sum_{a} l_{i}^{(a)} n_{a} + 1)} z^{n}
\eeq

 Here the mirror map  $t_{a}$ is is defined as $t_{a} = \frac{\partial_{a}
\omega_{0}}{\omega_{0}}$.

\subsection{Toric Geometry and GKZ System}
~~~~The generalized hypergeometric system was first introduced in
refs. \cite{GKZ}, and soon after gained a fast development in mirror
symmetry \cite{Batyrev:9310003, Batyrev:9307010, Yau:9308122,
Yau:9511001, Yau:9707003}. Define a mirror pair of
hypersurfaces $(X, X^{*})$ in two toric ambient spaces $(V, V^{*})$, respectively. The toric varieties $(V, V^{*})$ are
related to the fans $(\Sigma(\Delta), \Sigma(\Delta^{*}))$
induced by two dual polyhedron $(\Delta, \Delta^{*})$. The
defining polynomial for the hypersurface is defined as:
\beq \mathcal{P} = \sum_{i=0}^{p-1} a_{i} \prod_{k=1}^{4} X_{k}^{v_{i,
k}^{*}} \eeq Or we can write the above equation in another way \beq
\mathcal{P} = \sum_{i=0}^{p-1} a_{i} \prod_{v_{j}\in\Delta}
x_{j}^{<v_{j}, v_{i}^{*}> +1} \eeq
Here $a_{i}$ is complex parameter and $X_{k}$ are inhomogeneous coordinates on the open torus, $x_{i}$ is the homogeneous coordinates.

The general integral period is expressed as
\beq \Pi(a_{i}) = \frac{1}{(2\pi i)^{4}}\int_{|X_{k}|=1} \frac{1}{P} \prod_{k=1}^{4} \frac{dX_{k}}{X_{k}} \eeq
It is referred in \cite{Batyrev:9310003, Batyrev:9307010} that the period can be annihilated by a GKZ hypergeometric differential system
\beq \mathcal{D}_{l} \Pi(a) = 0~~~(l\in L)~~,~~ \mathcal{Z}_{i} \Pi(a) = 0~~~(i = 0, 1, \cdots, p) \eeq
the operators $\mathcal{D}_{l}$ and $\mathcal{Z}_{j}$ are expressed as
\beq \mathcal{D}_{l} = \prod_{l_{i}>0} (\frac{\partial}{\partial a_{i}})^{l_{i}} - \prod_{l_{j}<0} (\frac{\partial}{\partial a_{j}})^{-l_{j}}~~~~~(l\in L) \eeq
\beq \mathcal{Z}_{j} = \sum_{i=0}^{p} \bar{v}_{i, j}^{*} \theta_{a_{i}} - \beta_{j}~~~~~(j = 0, 1, \cdots, n) \eeq
The torus invariant algebraic coordinates $z_{a}$ in the large complex structure limit \cite{Alim:1010.0977}
\beq z_{a} = (-1)^{l_{0}^{a}} \prod_{j} a_{j}^{l_{j}^{a}} \eeq
where $l^{a}$ is the set of basic vectors which denote the generators of the Mori cone. Then, by $\theta_{a} = z_{a} \partial_{z_{a}}$, (2.16) changes into
\beq \mathcal{D}_{l} = \prod_{k=1}^{l_{0}} (\theta_{0} - k) \prod_{l_{i}>0} \prod_{k=0}^{l_{i}-1} (\theta_{i} - k) - (-)^{l_{0}} z_{a} \prod_{k=1}^{-l_{0}} (\theta_{0} - k) \prod_{l_{i}<0} \prod_{k=0}^{-l_{i}-1} \eeq
where $l$ is the linear combination of $l^{a}$. One can refer \cite{Alim:0909.1842, Beem:0909.2245, Yau:0910.4215} for more details.
The result to the GKZ system is described as
\beq \mathcal{B}_{\{l^{a}\}}(z_{a}; \rho_{a}) = \sum_{n_{1},\cdots,n_{N} \in Z_{0}^{+}} \frac{\Gamma(1 - \sum_{a} l_{0}^{a} (n_{a} + \rho_{a}))}{\prod_{i>0} \Gamma(1 + \sum_{a} l_{i}^{a} (n_{a} + \rho_{a}))} \prod_{a} z_{a}^{(n_{a} + \rho_{a})} \eeq

\section{Study of two compact non-fermat type  Calabi-Yau manifolds}
~~~~In this section we will calculate the superpotentials and disk invariants for two compact CY in the weighted projective space, respectively,  with the method mentioned in the section 2.

\subsection{Calabi-Yau hypersurface $X_{7}(1, 1, 1, 1, 3)$}
~~~~$X_{7}(1, 1, 1, 1, 3)$ is a hypersurface in the weighted projective space $\mathbf{P}^{4}(1, 1, 1, 1, 3)$. Let $X$ denotes this hypersurface, then, its mirror manifold $X^{*}$ is denoted by $ X^{*} = \hat{X}/H $, where $\hat{X}$ represents the CY 3-fold $X_{14}(1, 2, 2, 2, 7)$ and $H$ is defined by $(h_{i}^{j})=\frac{1}{7}(1, 0, 6, 0, 0)~,~\frac{1}{7}(1, 6, 0, 0, 0)$. So, $X_{7}(1, 1, 1, 1, 3)$ is isomorphic to $X_{14}(1, 2, 2, 2, 7)$, which had been checked in refs. \cite{Katz:9412117}.
The hypersurface $X_{7}(1, 1, 1, 1, 3)$ is defined as the zero locus of the polynomial $\mathcal{P}$.

\beq \mathcal{P} = x_{1}^{7} + x_{2}^{7} + x_{3}^{7} + x_{4}^{7} + x_{4} x_{5}^{2} \eeq

The weighted projected space $\mathbf{P}^{4}(1, 1, 1, 1, 3)$ as a toric variety, the vertices of its corresponding polyhedron $\Delta$ are as follows:
$$ v_{1} = (-1, -1, -1, 1)~,~v_{2} = (-1, -1, -1, -1)~,~v_{3} = (-1, -1, 1, 0)~, $$
$$ v_{4} = (6, -1, -1, -1)~,~v_{5} = (-1, 1, -1, 0)~,~v_{6} = (-1, 3, -1, -1)~, $$
$$ v_{7} = (-1, -1, 3, -1)~,~v_{8} = (0, -1, 3, -1)~,~v_{9} = (0, 3, -1, -1)~, $$
The vertices of the corresponding dual polyhedron $\Delta^{*}$ are
$$ v_{1}^{*} = (-1, -1, -1, -3)~,~v_{2}^{*} = (1, 0, 0, 0)~,~v_{3}^{*} = (0, 1, 0, 0)~, $$
$$ v_{4}^{*} = (0, 0, 1, 0)~,~v_{5}^{*} = (0, 0, 0, 1)~,~v_{6}^{*} = (0, 0, 0, -1)~, $$
There exists only one integral point denoted as $v_{0}^{*} = (0, 0, 0, 0)$ in $\Delta^{*}$.
For $\Delta^{*}$, the charge vector of the Mori cone is
$$ l^{(1)} = (-2, 0, 0, 0, 0, 1, 1)~~,~~l^{(2)} = (-1, 1, 1, 1, 1, 0, -3) $$

Consider the divisor \beq Q(D) = x_{3}^{7} + z_{3} x_{4}^{7} \eeq
at the critical point
$z_{3} = 1$.
Let \beq u_{1} = -\frac{z_{1}}{z_{3}}(1 - z_{3})^{2}~~~,~~~u_{2} = z_{2} \eeq
then according to (2.19) the GKZ system of the two-parameters family become
\beq \mathcal{D}_{1} = \tilde{\theta}_{1}(\tilde{\theta}_{1} - 3\tilde{\theta}_{2}) - (2\tilde{\theta}_{1} + \tilde{\theta}_{2})(2\tilde{\theta}_{1} + \tilde{\theta}_{2} - 1)z_{1} \eeq
\beq \mathcal{D}_{2} = \tilde{\theta}_{2}^{2}(7\tilde{\theta}_{2} - 2\tilde{\theta}_{1}) + 4\tilde{\theta}_{2}^{2}(2\tilde{\theta}_{1} + \tilde{\theta}_{2}-1)z_{1} - 7\prod_{i=1}^{3}(2\tilde{\theta}_{2} - \tilde{\theta}_{1} - i) \eeq
The solution to this GKZ system is written as
\beq \Pi_{1}(u_{1}, u_{2}) = \frac{c}{2}B_{\{\tilde{l}\}}(u_{1}, u_{2}, 0, \frac{1}{2}) \eeq
\beq \Pi_{2}(u_{1}, u_{2}) = \frac{c}{2}B_{\{\tilde{l}\}}(u_{1}, u_{2}, \frac{1}{2}, \frac{1}{2}) \eeq
At the critical point $z_{3}=1$, the on-shell superpotential satisfies $W_{C}^{+} = W_{C}^{-}$ according to the $Z_{2}$ symmetry. So the on-shell superpotential is described as
\beq W^{\pm}(z_{1}, z_{2}) = \frac{1}{2\pi i}\int_{\xi_{0}}^{\pm\sqrt{z_{3}}}\Pi(z_{1}, z_{2}, \xi^{2})\frac{d\xi}{\xi}|_{z_{3}=1} \eeq
For this model the on-shell superpotentials are expressed as
\beq W_{1}^{\pm} = \mp\frac{c}{8} \sum_{n_{1},n_{2}\geq0}\frac{\Gamma(2n_{1} + n_{2} +\frac{3}{2}) z_{1}^{n_{1}} z_{2}^{n_{2}+\frac{1}{2}}}{\Gamma(n_{1} +1) \Gamma(n_{1} - 3n_{2} - \frac{1}{2}) \Gamma(n_{2} + \frac{3}{2})^{4}} \eeq
\beq W_{2}^{\pm} = \mp\frac{c}{8} \sum_{n_{1},n_{2}\geq0}\frac{\Gamma(2n_{1} + n_{2} +\frac{5}{2}) z_{1}^{n_{1}+\frac{1}{2}} z_{2}^{n_{2}+\frac{1}{2}}}{\Gamma(n_{1} +\frac{3}{2}) \Gamma(n_{1} - 3n_{2}) \Gamma(n_{2} + \frac{3}{2})^{4}} \eeq
The flat coordinates in A- and B-model are connected via mirror map $t_{a} = \frac{\partial_{a}\omega_{0}}{\omega_{0}}$.
The mirror map is as follows:
$$ t_{1} = \log(z_{1}) + 2z_{1} + 3z_{1}^{2} + \frac{20}{3}z_{1}^{3} + 68z_{1}^{2}z_{2} - 10z_{1}z_{2} + 2z_{2} - 15z_{2}^{2} + 66z_{1}z_{2}^{2} + \frac{560}{3}z_{2}^{3} + o(z^{3}) $$
$$ t_{2} = \log(z_{2}) - 6z_{2} + 45z_{2}^{2} - 560z_{2}^{3} - 198z_{2}^{2}z_{1} + 30z_{2}z_{1} + 9z_{1} - 204z_{2}z_{1}^{2} + \frac{43}{2}z_{1}^{2} + 62z_{2}^{3} + o(z^{3}) $$
and the corresponding inverse mirror map is
$$ z_{1} = q_{1} - 2q_{1}^{2} + 3q_{1}^{3} - 2q_{1}q_{2} + 5q_{1}q_{2}^{2} + 36q_{1}^{2}q_{2} + o(q^{4}) $$
$$ z_{2} = q_{2} + 6q_{2}^{2} + 9q_{2}^{3} - 9q_{1}q_{2} - 120q_{1}q_{2}^{2} + 37q_{1}^{2}q_{2} + o(q^{4}) $$
According to (2.10) we can derive the Ooguri-Vafa invariants from the on-shell superpotentials. The results are listed in the following tables:

 \begin{table}[!h]
\def\temptablewidth{1.0\textwidth}
\begin{center}
\begin{tabular*}{\temptablewidth}{@{\extracolsep{\fill}}c|ccccccc}
$d_1\diagdown d_2 $&$1$     &$3$          &$5$      &$7$      &$9$      &$11$          \\\hline
$0$                     & $-2 $     &$2$                            &$-10$                       &$ 84 $                 &$-858$             &$13820$      \\
  $1$                     & $28$     &$-28$                         &$252 $    &$-2828$            & $36400$      &$-729130$       \\
   $2$                      &$70$      &$112$                   &$-2702$              &$42910 $              & $-714140$       &$17644120$     \\
   $3$                        & $0$      &$252$              &$16716$               &$-391580$      & $8645280$    &$-262033434$  \\
   $4$                          & $0$ &$6832$              &$-83286$    &$2543170$  &$-74112654$     &$2741674588$        \\
   $5$                           & $0 $     &$84364$         &$315980$         &$-13445796$          &$496350736$     &$-22527070394$    \\
    \end{tabular*}
       {\rule{\temptablewidth}{1pt}}
       \tabcolsep 0pt \caption{$n_{d_{1}, d_{2}}^{(1,+)}$}
\end{center}
\end{table}
\begin{table}[!h]
\def\temptablewidth{1.0\textwidth}
\begin{center}
\begin{tabular*}{\temptablewidth}{@{\extracolsep{\fill}}c|ccccccc}
$d_1\diagdown d_2 $&$1$     &$3$          &$5$      &$7$      &$9$      &$11$          \\\hline
$1$                     & $0 $     &$0$                            &$0$                       &$ 0 $                 &$0$             &$0$      \\
  $3$                     & $-70$     &$0$                         &$0 $    &$0$            & $0$      &$137970$       \\
   $5$                      &$-28$      &$56$                   &$-140$              &$896 $              & $-8008$       &$-5747938$     \\
   $7$                        & $2$      &$-792$              &$3164$               &$-27608$      & $315050$    &$108056554$  \\
   $9$                          & $0$     &$-38962$              &$-32424$    &$391104$  &$-5784408$     &$-1230725514$        \\
   $11$                           & $0$     &$-84364$         &$358288$         &$-3814860$          &$68819744$     &$10291361734$    \\
    \end{tabular*}
       {\rule{\temptablewidth}{1pt}}
       \tabcolsep 0pt \caption{$n_{d_{1}, d_{2}}^{(2,+)}$}
\end{center}
\end{table}

\subsection{Calabi-Yau hypersurface $X_{9}(1, 1, 2, 2, 3)$}
~~~~$X_{9}(1, 1, 2, 2, 3)$ is a hypersurface defined in the weighted projective space $\mathbf{P}^{4}(1, 1, 2, 2, 3)$. Let $X$ denotes the corresponding Calabi-Yau threefold, then, its mirror manifold $X^{*}$ is denoted by $ X^{*} = \hat{X}/H $, where $\hat{X}$ represents the CY 3-fold $X_{12}(1, 1, 3, 3, 4)$ and $H$ is defined by $(h_{i}^{j})=\frac{1}{9}(1, 8, 0, 0, 0)~,~\frac{1}{4}(1, 0, 3, 0, 0)$. So, $X_{9}(1, 1, 2, 2, 3)$ is isomorphic to $X_{12}(1, 1, 3, 3, 4)$.
The polyhedron $\Delta$ for this model has the vertices
$$ v_{1} = (-1, -1, -1, 2)~,~v_{2} = (-1, -1, -1, -1)~,~v_{3} = (-1, -1, 2, 0)~, $$
$$ v_{4} = (8, -1, -1, -1)~,~v_{5} = (-1, 2, -1, 0)~,~v_{6} = (-1, 3, -1, -1)~, $$
$$ v_{7} = (-1, -1, 3, -1)~,~v_{8} = (0, -1, 3, -1)~,~v_{9} = (0, 3, -1, -1)~, $$
then the dual polyhedron $\Delta^{*}$ are
$$ v_{1}^{*} = (-1, -2, -2, -3)~,~v_{2}^{*} = (1, 0, 0, 0)~,~v_{3}^{*} = (0, 1, 0, 0)~, $$
$$ v_{4}^{*} = (0, 0, 1, 0)~,~v_{5}^{*} = (0, 0, 0, 1)~,~v_{6}^{*} = (0, -1, -1, -1)~, $$
There exist no points inside the polyhedron $\Delta^{*}$ but the original point $v_{0}^{*} = (0, 0, 0, 0)$.
We define the hypersurface $X_{9}(1, 1, 2, 2, 3)$ as the zero locus of the above polynomial $\mathcal{P}$.
\beq \mathcal{P} = x_{1}^{9} + x_{2}^{9} + x_{1} x_{3}^{4} + x_{2} x_{4}^{4} + x_{5}^{3} \eeq
$$ l^{(1)} = (-3, -1, -1, 1, 1, 0, 3)~~,~~l^{(2)} = (-1, 1, 1, 0, 0, 1, -2) $$
are the generators of the Mori cone related to this model.

Consider the divisor \beq Q(D) = x_{1}^{9} + z_{3} x_{2}^{9} \eeq
at the critical point
$z_{3} = 1$.
Let \beq u_{1} = -\frac{z_{1}z_{3}}{(1 - z_{3})^{2}}~~~,~~~u_{2} = -\frac{z_{2}}{z_{3}}(1 - z_{3})^{2} \eeq
according to (2.16) the GKZ system for the two-parameters family become
\beq \mathcal{D}_{1} = \tilde{\theta}_{2}(\tilde{\theta}_{2} - \tilde{\theta}_{1})^{2} - (3\tilde{\theta}_{1} + \tilde{\theta}_{2})(3\tilde{\theta}_{1} - 2\tilde{\theta}_{2} + 2)(3\tilde{\theta}_{1} - 2\tilde{\theta}_{2} + 1)z_{2} \eeq
$$ \mathcal{D}_{2} = (\tilde{\theta}_{2} - \tilde{\theta}_{1})^{2} - (\tilde{\theta}_{2} - \tilde{\theta}_{1})(3\tilde{\theta}_{1} - 2\tilde{\theta}_{2}) + 4\tilde{\theta}_{1}(3\tilde{\theta}_{1} - 2\tilde{\theta}_{2}) + 3z_{1}(3\tilde{\theta}_{1} - 2\tilde{\theta}_{2})(3\tilde{\theta}_{1} - 2\tilde{\theta}_{2} - 1)$$
$$ - 48z_{1}z_{2}(3\tilde{\theta}_{1} + \tilde{\theta}_{2} + 1)(3\tilde{\theta}_{1} + \tilde{\theta}_{2} + 2) - 48z_{1}z_{2}(3\tilde{\theta}_{1} + \tilde{\theta}_{2} + 1)(3\tilde{\theta}_{1} - 2\tilde{\theta}_{2}) - 16z_{1}(\tilde{\theta}_{2} - \tilde{\theta}_{1})^{2} $$
The solution to this GKZ system are written as \beq
\Pi_{1}(u_{1}, u_{2}) = \frac{c}{2}B_{\{\tilde{l}\}}(u_{1}, u_{2},
\frac{1}{2}, 0) \eeq \beq \Pi_{2}(u_{1}, u_{2}) =
\frac{c}{2}B_{\{\tilde{l}\}}(u_{1}, u_{2}, \frac{1}{2}, \frac{1}{2})
\eeq Similarly, in this model the on-shell superpotentials satisfy
$W_{C}^{+} = W_{C}^{-}$ according to the $Z_{2}$ symmetry. So
the superpotentials is described as \beq W^{\pm}(z_{1}, z_{2})
= \frac{1}{2\pi i}\int_{\xi_{0}}^{\pm\sqrt{z_{3}}}\Pi(z_{1}, z_{2},
\xi^{2})\frac{d\xi}{\xi}|_{z_{3}=1} \eeq At the critical point $z_{3}=1$, on-shell
superpotentials are expressed as \beq W_{1}^{\pm} = \mp\frac{c}{8}
\sum_{n_{1},n_{2}\geq0}\frac{\Gamma(3n_{1} + n_{2} + \frac{5}{2})
z_{1}^{n_{1} + \frac{1}{2}} z_{2}^{n_{2}}}{\Gamma(-n_{1} + n_{2} +
\frac{1}{2})^{2} \Gamma(3n_{1} - 2n_{2} + \frac{5}{2})
\Gamma(n_{2}+1) \Gamma(n_{1} + \frac{3}{2})^{2}} \eeq \beq
W_{2}^{\pm} = \mp\frac{c}{8}
\sum_{n_{1},n_{2}\geq0}\frac{\Gamma(3n_{1} + n_{2} + 3) z_{1}^{n_{1}
+ \frac{1}{2}} z_{2}^{n_{2} + \frac{1}{2}}}{\Gamma(-n_{1} + n_{2} +
1)^{2} \Gamma(3n_{1} - 2n_{2} + \frac{3}{2}) \Gamma(n_{2}+
\frac{3}{2}) \Gamma(n_{1} + \frac{3}{2})^{2}} \eeq
the mirror maps are
$$ t_{1} = \log(z_{1}) + 30z_{1}z_{2} - 3z_{2}  + 252z_{1}z_{2}^{2} -10z_{2}^{3} + 927z_{1}^{2}z_{2}^{2} + 288z_{1}z{2}^{3} - \frac{105}{4}z_{2}^{4} +  o(z^{4}) $$
$$ t_{2} = \log(z_{2}) + 2z_{2} + 3z_{2}^{2} + \frac{20}{3}z_{2}^{3} + 74z_{1}z_{2} - 168z_{1}z_{2}^{2} - 192z_{1}z_{2}^{3} + 8853z_{1}^{2}z_{2}^{2} + \frac{35}{2}z_{2}^{4} + o(z^{4}) $$
and the corresponding inverse mirror maps are
$$ z_{1} = q_{1} + 3q_{1}q_{2} + 3q_{1}q_{2}^{2} + q_{1}q_{2}^{3} - 30q_{1}^{2}q_{2} -594q_{1}^{2}q_{2}^{2} + o(q^{4}) $$
$$ z_{2} = q_{2} - 2q_{2}^{2} + 3q_{2}^{3} - 4q_{2}^{4} -74q_{1}q_{2}^{2} + 390q_{1}q_{2}^{3} + o(q^{4}) $$
Analogous to computing the disk invariants of the $X_{7}(1, 1, 1, 1, 3)$, we have
the results listed in the following tables:

\begin{table}[!h]
\def\temptablewidth{1.0\textwidth}
\begin{center}
\begin{tabular*}{\temptablewidth}{@{\extracolsep{\fill}}c|ccccccc}
$d_1\diagdown d_2 $&$0$     &$1$          &$2$      &$3$      &$4$      &$5$          \\\hline
$1$                     & $2 $     &$18$                            &$2$                       &$ 2 $                 &$2$             &$2$      \\
  $3$                     & $0$     &$0$                         &$1584 $    &$-710$            & $-626$      &$-616$       \\
   $5$                      &$0$      &$0$                   &$0$              &$38018 $              & $208244$       &$49382$     \\
   $7$                        & $0$      &$0$              &$0$               &$0$      & $1745190$    &$111219514$  \\
   $9$                          & $0$ &$0$              &$0$    &$0$  &$0$     &$90081018$        \\
   $11$                           & $0 $     &$0$         &$0$         &$0$          &$0$     &$-94519326$    \\
    \end{tabular*}
       {\rule{\temptablewidth}{1pt}}
       \tabcolsep 0pt \caption{$n_{d_{1}, d_{2}}^{(1,+)}$}
\end{center}
\end{table}
\begin{table}[!h]
\def\temptablewidth{1.0\textwidth}
\begin{center}
\begin{tabular*}{\temptablewidth}{@{\extracolsep{\fill}}c|ccccccc}
$d_1\diagdown d_2 $&$1$     &$3$          &$5$      &$7$      &$9$      &$11$          \\\hline
$1$                     & $16 $     &$0$                            &$-2$                       &$ -2 $                 &$-2$             &$46$      \\
  $3$                     & $0$      &$-80 $    &$2208$            & $608$      &$584$   &$26332$       \\
   $5$                      &$0$      &$0$                   &$720$              &$158784 $              & $-4120$       &$3039542$     \\
   $7$                        & $0$      &$0$              &$0$               &$-8848$      & $15904928$    &$94908448$  \\
   $9$                          & $0$ &$0$              &$0$    &$0$  &$126608$     &$3473058320$        \\
   $11$                           & $0 $     &$0$         &$0$         &$0$          &$0$     &$-434510208$    \\
    \end{tabular*}
       {\rule{\temptablewidth}{1pt}}
       \tabcolsep 0pt \caption{$n_{d_{1}, d_{2}}^{(2,+)}$}
\end{center}
\end{table}
\section{Conclusion}
~~~~The D-brane superpotential plays a crucial role in both physics and mathematics. From the physical point of view,
they determine the vacuum of the low energy effective theory. From the A-model worldsheet viewpoint, it is  the generating functional of
the Ooguri-Vafa Invariants of Calabi-Yau manifold and the submanifold on which is wrapped by the D-branes in the A-model.
these Ooguri-Vafa Invariants are closely related to the number of the BPS states.
 From the mirror geometric viewpoint, it is the integral period which is the solution to the generalized GKZ system.
 It is very hard to calculate directly the D-brane superpotential for compact Calabi-Yau manifold in A-model because these superpotentials
 are non-perturbative in essential, and  are impossibly obtained from the  perturbative  or localization way which is important methods
 to compute the D-brane superpotential in non-compact Calabi-Yau manifold in A-model. A effective approach to obtain the D-brane superpotential is
  by using the blown-up geometry of target space along the submanifold  wrapped by the D-branes \cite{B, ee}. The  alternative approach to compute
  the superpotential  of the D-brane in compact Calabi-Yau manifold in A-model is via the algebraic geometric method and mirror symmetry.

  In this paper, we extend the generalized GKZ system in a Fermat Calabi-Yau threefolds to the compact non-Fermat Calabi-Yau threefolds
 which are less study so far in contrast to the Fermat type Calabi-Yau threefolds. we first construct the generalized GKZ system
 for the compact non-fermat type Calabi-Yau manifolds, then  work out the corresponding D-brane superpotential  in the mirror B-model
 by the algebraic geometric method. The superpotential in the A-model is obtained accroding to mirror symmetry. Finally
 the Ooguri-Vafa invariants are extracted from  the A-model superpotential.

     These superpotential have potential phenomenological applications. Furthermore, acoording to the type II string/M-theory/F-theory duality,
    in the weak decoupling limit $g_s\rightarrow0$,  these superpotentials of Type II string  give  the Gukov-Vafa-Witten superpotentials $\mathcal{W}_{GVW}$ of
   F-theory compactified on the dual fourfold.  On the other hand, since there is not a systematic mathematical method to compute them by now, after all,
   it is difficult to get, from other approach, those Ooguri-Vafa Invariants predicted in this paper.
Those Ooguri-Vafa Invariants  provide some concrete data which could potentially be checked by an independent mathematical calculation.

\section*{Acknowledgments}
~~~~The work is supported by the NSFC (11075204) and NSFC(11475178).


\addcontentsline{toc}{section}{References}

\begingroup\raggedright\endgroup

\end{document}